# Two-dimensional Clay Channels for Tunable Nanofluidic Memristor

Sangeeta Yadav,[1,2,3,†] Raj Kumar Gogoi,[1,2,†] Aziz Lokhandwala,[1,2] Siddhi Vinayak Pandey,[1,2] Ankit Bhardwaj,[1,2] Sunando DasGupta,[4] Boya Radha[1,2,5,*]

[1]Department of Physics and Astronomy, The University of Manchester, Manchester M13 9PL, U.K.
[2]National Graphene Institute, The University of Manchester, Manchester M13 9PL, U.K.
[3]Advanced Technology Development Centre, Indian Institute of Technology Kharagpur, Kharagpur 721302, India
[4]Department of Chemical Engineering, Indian Institute of Technology Kharagpur, Kharagpur 721302, India
[5]Photon Science Institute, University of Manchester, Manchester M13 9PL, U.K.
*Corresponding email: radha.boya@manchester.ac.uk
[†]Contributed equally.

## Abstract

Dynamic reconfiguration of charge carriers in confined ion-channels under electrical stimulation produces memory effects, where the internal resistance depends on history of the electric field. Vermiculite nanofluidic devices harness this effect to store and process information within a single component. We report switching between distinct memory loops by tuning ion transport pathways, governed by asymmetrical device architecture and intrinsic surface-charge. Polarity-dependent memory switching between crossing-1 and crossing-2 loops is achieved solely by altering electrode configurations, without modifying electrolyte, channel surface chemistry or device structure: providing mechanistic insights into ionic memristors through a straightforward, experimental strategy. The memristive characteristics are demonstrated in both *in-plane* and *out-of-plane* channel configurations with channel lengths spanning from centimeters to micrometers length scales using re-stacked vermiculite membranes and further investigated for miniaturization with devices of nanometer scale channel lengths, fabricated via ultramicrotomy method. Furthermore, we demonstrate neuromorphic functionalities, including synaptic potentiation-depression and programmable memory retention, highlighting potential for bio-inspired computing systems. Cost-effective and scalable fabrication of solution processed vermiculite membrane memristors is attractive for sustainable integration of nanofluidic memristors for neuromorphic computing applications.

## Introduction

The human brain, with a neural network of approximately $10^{11}$ neurons connected with $10^{15}$ synapses, facilitates complex cognitive functions like learning, memory and sensory processing with exceptional energy efficiency.[1] This led the quest to develop neuromorphic devices for efficient information storage, transmission and processing.[2,3] Consequently, *memristor*, first theorized by Leon O. Chua[4] in 1971 and experimentally realized as a solid-state physical device using Ti/$TiO_x$ nanojunctions by Stanley et al.[5] in 2008 have garnered significant attention for their ability to both store and compute information simultaneously. However, despite the development and exploration of numerous solid-state memristor devices based primarily on electronic transport mechanisms, emulating the full complexity of brain-like computation remains a formidable challenge.[6,7] The biological synapses operate *via* ionic fluxes, electrochemical and chemical signalling, rather than solely through electron conduction.[8,9] This fundamental mismatch in the mode of signal transduction limits the fidelity with which solid-state devices can replicate key synaptic functions such as plasticity, adaptation, and energy-efficient signal processing. To bridge this gap, there is growing interest in neuromorphic devices that harness ion transport, particularly within soft matter,[10] electrolytes,[11] and nanofluidic systems,[12-16] which can modify their conductance or current based on voltage history.[10,17,18] Recent advances in fabrication techniques[19-22] and development in fluidic or ionic memristors show promise in mimicking synaptic behaviours, including long- and short-term plasticity, through voltage-induced modulation of ionic currents and interfacial charge distributions as shown in 2D nanochannels.[12]

In this study, we used vermiculite clay, which is a widely available natural 2D material, to design ionic memristor by utilizing the 2D channels present in lamellar restacked membrane of 2D nanosheets. The lamellar channels formed in restacked clay membranes provide nanoconfined pathways for ion transport, enabling memristive behaviour driven by voltage history. Intriguingly, certain hypothesis on the origin of life suggest that clay minerals may have played a role in prebiotic chemistry,[23] facilitating the organization of simple molecules on their surfaces. Inspired by this notion of primitive molecular templating, we explore clay-based nanofluidic systems as platforms for bio-inspired ionic memory, where information is encoded in conductance states that evolve. Moreover, natural clay-based functional materials provide sustainable advantages due to their natural abundance, environmental compatibility, low-cost and scalable fabrication[24] and has inspired the modern research to strategically design artificially intelligent functional materials such as critical separations,[25] shape-memory actuators,[26,27] self-healing materials[27] and memristors.[28] Herein, we successfully demonstrate the tunability of vermiculite (VM) memristor with two memory types governed by the interplay of ionic movement and structural asymmetry created in the memristor design. We further show the electrolytic concentration dependent memristive nature of the vermiculite memristors in different channel configurations and its potential in neuromorphic computing applications.

**Results and Discussion**

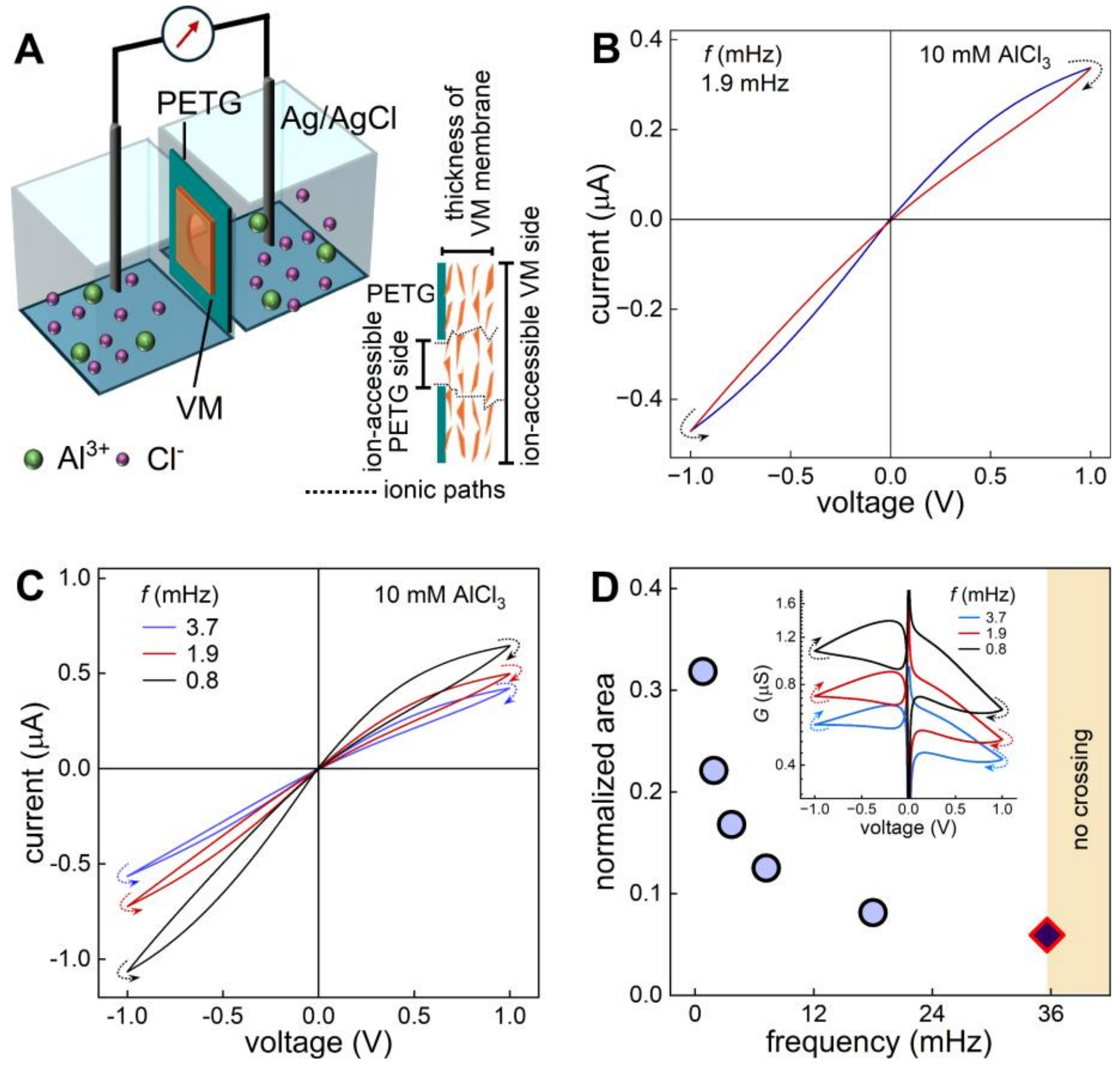

**Fig. 1: Vermiculite (VM) nanofluidic memristor.** (**A**) Schematic representing the experimental set-up with VM membrane (channel height, 6.3 Å) mounted on PETG with a circular hole of diameter 4 mm for ion transport measurements. The inset shows the path for ion transport across the membrane. (**B**) The voltage dependent ionic current (*IV*) characteristics of vermiculite nanofluidic device at 1.9 mHz frequency. (**C**) Frequency dependent (0.8, 1.9, and 3.7 mHz) *IV* characteristics of a vermiculite nanofluidic device. (**D**) Area under *IV* curves at different frequency (corresponding *IV*s are in Fig. 1C and Fig. S11). Inset in (**D**) shows the voltage-dependent-conductance (*GV*) curves of the *IV*s at 3.7, 1.9 and 0.8 mHz. The electrolyte used in Fig. 1 is 10 mM $AlCl_3$. Device V2 is used in (**B**) and V1 in (**C**) and (**D**), details in the Table S1.

We prepare the 2D VM channels, by restacking exfoliated 2D layers of vermiculite crystal into a lamellar membrane (Fig. S1),[29] with height ~ 6.3 Å (described in *section 2* in the SI). A rectangular strip (10 mm × 10 mm) of a membrane (thickness ~18 μm) is secured over a circular hole (4 mm diameter) of polyethylene terephthalate glycol (PETG) substrate using polydimethylsiloxane (PDMS), ensuring the channels are in an *out-of-plane* configuration (Fig. S7). The exposed area of the membrane towards the PETG hole side is ~12.57 $mm^2$ and the other side is ~100 $mm^2$, inducing a surface area asymmetry. To perform the ion-transport through these Å-scale channels, the device is mounted in between two reservoirs filled with the desired electrolyte. Home-made Ag/AgCl electrodes connected to a Keithley sourcemeter are inserted into the reservoirs (Fig. 1A). Different voltage ranges (from ±100 mV to ±1 V, Fig. S10) were used to record the voltage-dependent-current (*IV*) characteristics of the VM channels with 10 mM $AlCl_3$ as electrolyte, fixing the scan rate of 5 $mV.s^{-1}$. The *IV*s are non-linear with different ionic currents for a given voltage in the forward and reverse sweeps, forming a hysteresis loop that self-crosses at zero bias, Fig. 1B. This self-crossing behaviour is a hallmark of bipolar memristive systems, where conductance switching is governed by voltages of opposite polarity.[12] Systematic variation of the applied voltage and frequency (*f*) reveals precise tunability of the memristive response: both hysteresis and non-linearity decrease as the voltage range is reduced and nearly disappear around ±100 mV (Fig. S10). Likewise, increasing the *f* above ~35.6 mHz suppresses the hysteresis, whereas lower *f* enhances it (Fig. 1D shows the *f* dependent area, with corresponding *IV*s in Figs. 1C and S11). Similar *IV* hysteresis loops (Figs. 1C, S13-S15) were observed across multiple devices (device details: Table S1) fabricated from multiple VM membranes, confirming the reproducibility of the bipolar memristive behaviour of our VM devices at varied frequencies. The conductance (*G*) states are polarity-dependent confirming the bipolar memristive characteristics (Fig. 1D inset).

The observed hysteresis loops can be categorized into two types: crossing-1 curves which are characterized by reduction in *G* under positive bias, followed by an increase in *G* at negative bias; the crossing-2 curves exhibit the opposite behaviour. Our VM ionic memristor facilitates switching of the memory types from crossing-1 to crossing-2 (Fig. 2B, C) merely by interchanging the position of the working electrode (*WE*) and the counter electrode (*CE*) in the electrochemical setup, even under identical experimental conditions (schematic of the experimental condition in Fig. 2A). This inversion modulates the direction of ionic asymmetry within the device, arising from the combination of surface-charge-driven ion selectivity and structural asymmetry between the membrane and PETG opening. While the area asymmetry between the membrane and PETG hole sides leads to different number of accessible ion transport pathways on both sides of the membrane and hence introduces a directional resistance at ion entry and exit,[30,31] the negatively charged surface of VM channels offers cation dominated ion transport.[29] The entry of the charge carriers into the device can be tuned by the position of the *WE* as well as with the polarity of applied electric field.

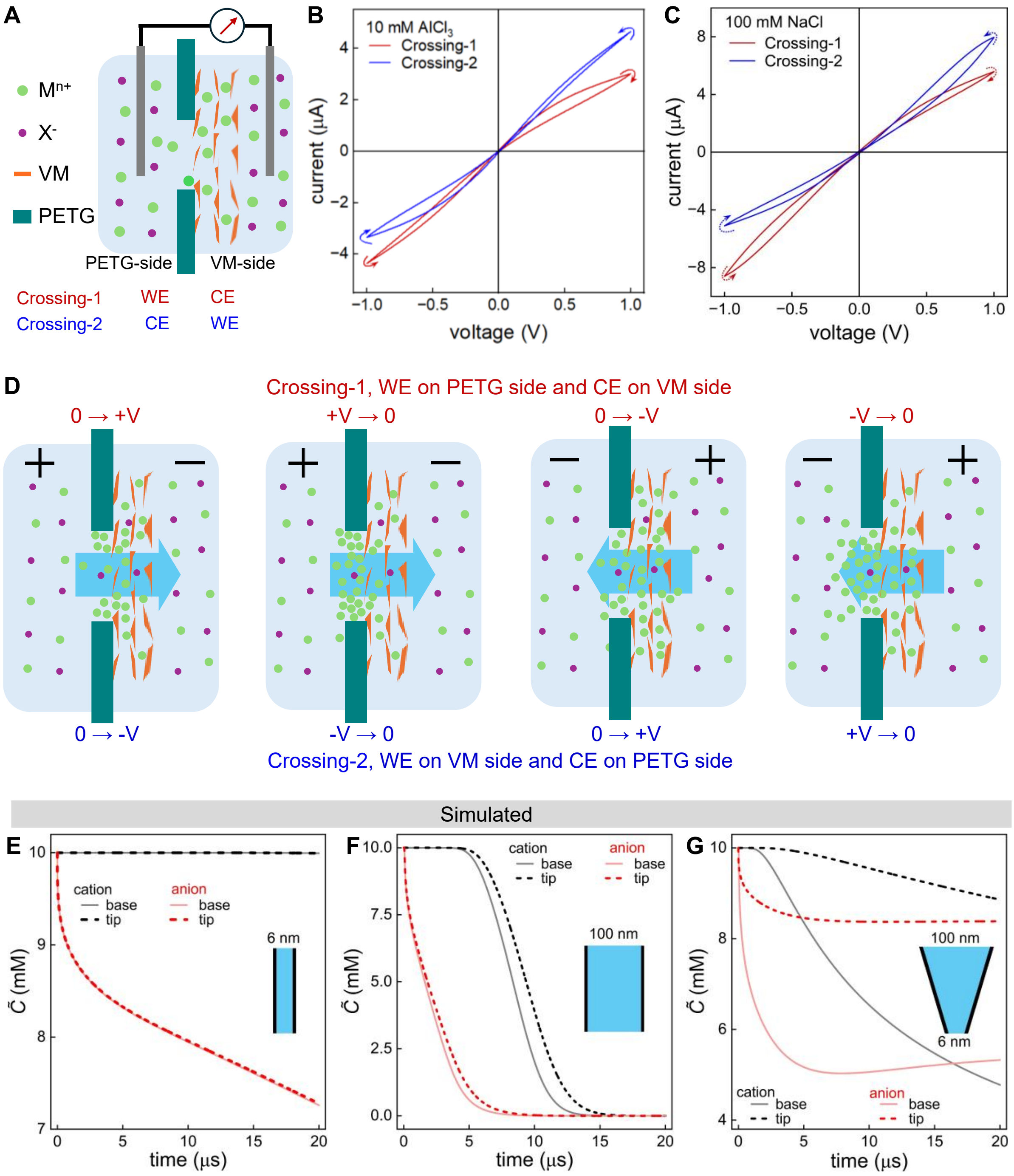

**Fig. 2. Switching of bipolar memory (crossing-1 to crossing-2):** (**A**) Schematic showing the electrode configuration for the *IV* measurements (where *WE* and *CE* refer to working and counter electrodes, respectively). The switching of the memory loop types for (**B**) 10 mM $AlCl_3$ and (**C**) 100 mM NaCl as electrolytes (V3 is used in (**B**) and V2 in (**C**), details in the Table S1 and frequency used for the *IV* measurement is of 0.8 mHz. (**D**) Schematic representation of the mechanism of the memristive behaviour for crossing-1 and crossing-2. Simulation of time dependent average concentration variation ($\tilde{C}$) of cations and anions (estimated using equation S6 and S7, respectively) near the tip and base at +1V (*WE* and *CE* on the base and tip side, respectively) for cylindrical nanochannel with a diameter of (**E**) ~6 nm and (**F**) ~100 nm and (**G**) for a conical nanochannel (tip and base diameters of ~6 nm and ~100 nm, respectively). The surface-charge-density of channel wall in (**E**), (**F**) and (**G**) is assumed to be -10 mC.m$^{-2}$ and the electrolyte concentration is assumed as 10 mM.

To illustrate the mechanism underlying the crossing-1 loop, Fig 2B, consider a positive bias applied to the PETG side (Fig. 2D, 0 → +V, dark red colour). Here, cations ($M^{n+}$) drift from the PETG opening into the membrane, where they encounter high access resistance due to incoming ionic flux, leading to accumulation and crowding at the channel entrance (decreasing the rate of ionic flow inside the channels). This crowding impedes further ion transport and hence reducing the rate of increase in current with increase in applied voltage. As the voltage is subsequently swept back (Fig. 2D, +V → 0), the weakened electric field is insufficient to overcome the crowding barrier, further suppressing the current. In contrast, when a negative bias is applied to the PETG side (Fig. 2D, 0 → –V), cations migrate from the VM side (increasing the ionic concentration inside the channels). Upon decreasing the voltage (Fig. 2D, –V → 0), this accumulation inside the channels disperses, enhancing the reverse current. Reversing the electrode configuration flips the direction of ionic drift and accumulation, inverting the memory loop from crossing-1 to crossing-2 (Fig. 2D). The switching of the memristive loops from crossing-1 to crossing-2 is observed in multiple devices for different voltage ranges, frequency and electrolyte concentrations (Figs. S17, S18 and 2B) as well as in another electrolyte (100 mM NaCl, Fig. 2C).

To support our proposed mechanism, we simulated the effect of the geometrical constraints on the ionic concentrations in the entry and exit of our device. The presence of memristive behaviour in devices with asymmetric geometry,[32] primarily relies on the ion concentration inside the channels ($C_{\mathrm{ch}}$) or conductance ($G$) as a function of accumulation/depletion, leading to relaxation of concentration differences near the entrances, as shown in Fig. 2D. The electrophoretic transport which occurs near the entrances is therefore due to the concentration difference present at a given instant between channel and the reservoir, given by $\alpha \cdot V$. This concentration modulation can be expressed as:

$$\frac{\mathrm{d}(C_{\mathrm{ch}}-C_{\mathrm{bulk}})}{\mathrm{dt}} = \alpha \cdot V - \frac{C_{\mathrm{ch}}-C_{\mathrm{bulk}}}{\tau(V)} - \frac{C_{\mathrm{ch}}}{\tau_{\mathrm{ads}}} + \frac{C_{\mathrm{ads}}}{\tau_{\mathrm{des}}} \quad (1)$$

$$\frac{\mathrm{d}C_{\mathrm{ads}}}{\mathrm{dt}} = \frac{C_{\mathrm{ch}}}{\tau_{\mathrm{ads}}} - \frac{C_{\mathrm{ads}}}{\tau_{\mathrm{des}}} \quad (2)$$

Where $V$ is applied voltage, $\alpha$ is a coefficient for voltage induced concentration difference between reservoir and the channel, $\tau(V)$ is relaxation time constants for electro-osmosis/diffusion and $\tau_{\mathrm{ads}}$ and $\tau_{\mathrm{des}}$ are adsorption and desorption rates for surface-charge governed ion-transport within the channel. For a sufficiently large reservoir, $\frac{dC_{\mathrm{bulk}}}{dt} = 0$. Thus, equation 1 can be expressed as:

$$\frac{dC_{\mathrm{ch}}}{dt} = \alpha \cdot V - \frac{C_{\mathrm{ch}} - C_{\mathrm{bulk}}}{\tau(V)} - \frac{C_{\mathrm{ch}}}{\tau_{\mathrm{ads}}} + \frac{C_{\mathrm{ads}}}{\tau_{\mathrm{des}}} \tag{3}$$

The $\alpha$ in equation (3) at a given instant ($dt$) is given by:

$$\alpha \cdot dt = \frac{G}{2 \cdot N_{\mathrm{A}} \cdot e} \cdot \left( \frac{1}{1 + D_{\mathrm{u}}^{\mathrm{PETG}}} - \frac{1}{1 + D_{\mathrm{u}}^{\mathrm{VM}}} \right) \cdot dt \tag{4}$$

The Dukhin number ($D_{\mathrm{u}}$) on PETG hole and VM side of our device in equation (4) is defined as $D_{\mathrm{u}}^{\mathrm{PETG}}$ and $D_{\mathrm{u}}^{\mathrm{VM}}$, respectively (in detailed, SI section 15).

Additionally, the hysteretic effects at high frequencies due to contribution from electro-osmosis/diffusion are not observed in the presented VM channels (Fig. 1D). In our channels, although the high number of ionic pathways leads to reduced effective diffusion length per path, the overall relaxation time increases causing large variations in local electric fields, leading to correlated ionic transport, and thereby hysteretic effects.

To qualitatively study the origin for accumulation/depletion processes due to the geometry (i.e. the $\alpha$), and eventually the origin of distinct $D_{\mathrm{u}}$ near entrance and exits, we performed 2D-axisymmetric Poisson-Nernst Planck (*PNP*) mean field-simulations for three different geometries; two cylindrical nanochannels each with diameter of ~6 nm and ~100 nm, and a conical nanochannel with tip diameter of ~6 nm and base diameter of ~100 nm to qualitatively study electric-field screening effects. To investigate the relaxation mediated by α in equation 3, we initialized $C_{\mathrm{ch}} = C_{\mathrm{bulk}}$ for all three geometries and probed the time dependent average concentration variation $\tilde{C}$ of individual ions along the *axis-of-symmetry* (equations S6 and S7), from the ends of the geometry to the bulk reservoir, mediated either by diffusion or surface charge, (Fig. 2B, C). Keeping the voltages in each of the geometries as ±1 V between the two reservoirs, we examine the concentration change (Fig. 2E-G for +1V). The relaxation times of the ions (both cations and anions) are dependent on the geometry of the channels; the decay rate is notably distinct between the entrance and exit of conical nanochannel. Qualitatively this suggests that geometrical asymmetry intrinsic with VM devices is likely the cause for the crowding/depletion processes[33] and the eventual hysteretic *IV* characteristics observed.

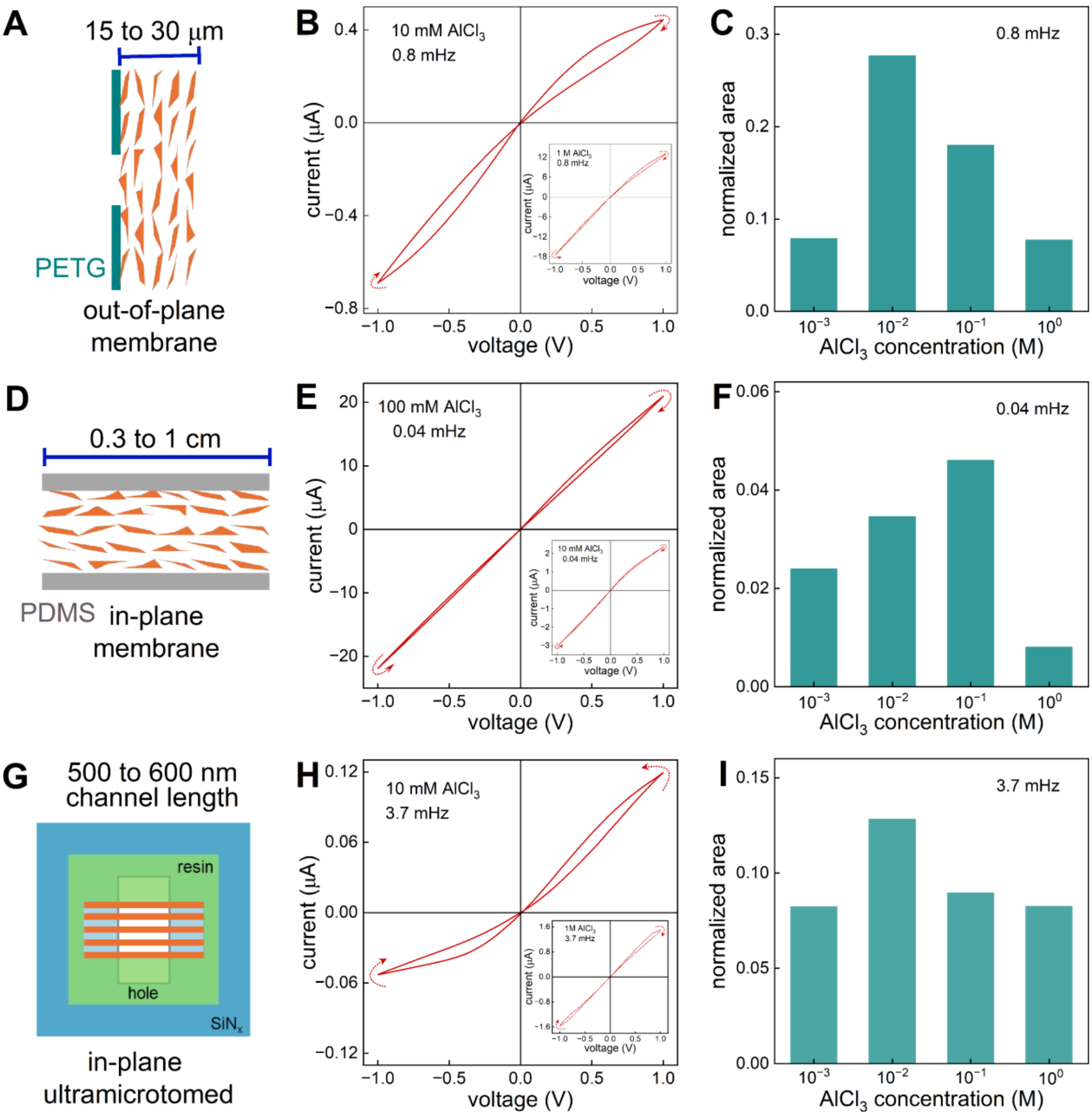

**Fig. 3. Memory in VM channels of different length scales:** Schematic showing the VM for (**A**) *out-of-plane* channels with $l$ in μm and (**D**) *in-plane* channels with $l$ in cm of VM membranes and (**G**) ultramicrotomed pristine in plane channels with $l < 1$ μm. (**B**) The *IV* characteristics for the device corresponding to **(A)** for 10 mM $AlCl_3$ and for 1M $AlCl_3$ in the right bottom inset at 0.8 mHz frequency. (**E**) The *IV* characteristics of the device corresponding to (**D**) for 100 mM $AlCl_3$ and for 10 mM $AlCl_3$ in the right bottom inset at 0.04 mHz frequency. (**F**) The enclosed area of the memristive loops at different electrolyte concentrations corresponding to the device configuration in (**D**). **(H)** The *IV* characteristics of the device corresponding to (**G**) for 10 mM $AlCl_3$ and for 1 M $AlCl_3$ in the right bottom inset at 3.7 mHz frequency. The normalized area of the memristive loops in (**C**), (**F**), (**I**) at different electrolyte concentrations corresponding to the channel configuration in (**A**), (**D**), (**G**) respectively. Device V2 is used in (**B**) and (**C**) and $D_{M1}$ is used in (**H**) and (**I**) (details in Fig. S31). For (**E**) and (**F**) membrane dimensions are 10 mm (length) × 5 mm (width) × 22 μm (thickness).

Moreover, the memristive behaviour was investigated for both the *out-of-plane* and *in-plane* ionic pathways of the VM channels with different concentration of $AlCl_3$ electrolyte, which exhibit $C$ and $f$ dependent memristive loops. In *out-of-plane* configuration, hysteresis is prominent at 10 mM *C*. The hysteresis curve in this configuration is obtained due to the asymmetry in the ion movement (termed ionic current rectification, *ICR*). At low *C*, the geometrical asymmetry drives the *ICR* of charged channels, but increasing *C* reduces the *ICR* due to the screening of the electrical double layers, [30,31,34] and thus decreases the memristive loop (Fig. 3B,C and S12). The symmetric *in-plane* VM membranes exhibit memristive behaviour only at low *f* and with minimal hysteresis (Fig. 3E) This can be attributed to longer channel length, in centimetre range, resulting from lateral transport across stacked 2D nanosheets. Since the exfoliated flakes have lateral dimensions of a few microns, centimeter-scale membranes expose a large number of edge sites, thereby increasing the effective surface-charge-density.[35] Additionally, the inherent geometric irregularities of the flakes may induce asymmetry in surface charge distribution at the two device ends. At low frequencies, ions have sufficient time to interact with these surface charges through adsorption–desorption processes, which contribute to complex ionic dynamics and enhanced hysteresis loop area. In contrast, the asymmetric *in-plane* membrane device with higher hysteresis (Fig. S30B).

To test our VM devices at sub-micrometer length-scales, we integrated VM flakes consisting of natural channels on $SiN_x$ chip and fabricated the devices with much smaller channel lengths of 500 to 600 nm with asymmetrical entry-exits *via* ultramicrotomy method,[36] The surface charge governed ionic conductivity for these miniaturized devices leads to a plateau for $C < 1$ mM, but for $C \geq 1$ mM, the conductivity increased linearly with *C* indicating the concentration dependent ionic transport regime.[36] Notably, these devices also exhibit bipolar memristive behaviour (Fig. 3H, S31E, F and S32), with a prominent hysteresis area at 10 mM $AlCl_3$ (Fig. S31), which decreases with either an increase or decrease in concentration. The similarity in the *IV*s behaviour of these nano-scale devices with the *out-of-plane* VM devices indicating that the memory effect is robust to miniaturization.

The hysteresis is prominent at 10 mM and 100 mM electrolyte concentration in all the device configurations (Fig. 3B, S12B, 3E, 3C, F, H and I). At low electrolyte concentrations (Fig. S12A, S29A, 3C, F and I), smaller voltage induced concentration difference ($\alpha$) due to low *G* combined with smaller surface charge regulation[37] will not vary the $C_{ch}$ significantly (equation 3 and 4) and hence leading to low hysteresis. However, in the intermediate concentration regime, the concentration dependence of surface charge terms in equation 3 dominates against concentration gradient term leading to significant hysteretic effects. Furthermore, at high concentrations, since all channels are occupied at entry and exits, $D_{\mathrm{u}}^{\mathrm{PETG}} \sim D_{\mathrm{u}}^{\mathrm{VM}}$, resulting in $\left(\frac{1}{1+D_u^{\mathrm{PETG}}}-\frac{1}{1+D_u^{\mathrm{VM}}}\right) \to 0$, and hence the hysteretic effects related to changing $\frac{dC_{\mathrm{ch}}}{dt}$ are diminished. Additionally, at higher concentrations, $C_{\mathrm{ch}} \sim C_{\mathrm{bulk}}$ leads to the merely absence of relaxation mechanisms such as diffusion, surface charge, and electro-osmosis. For symmetrical devices, although the Dukhin asymmetry due to geometry is absent, surface charge regulated processes may arises from the comparatively longer *l*, which exposes a larger number of flake edge sites within the device and consequently increases effective surface charge density[35] leading to the observed hysteresis effect (Fig. 3E). In asymmetric *in-plane* devices the presence of Dukhin's asymmetry and the surface charge regulated processes leads to the enhanced memristive behaviour (Fig. S30B). However, by tuning the flakes to smaller lateral flake sizes (Fig. S30C) the hysteretic behaviour can be enhanced further even in symmetric *in-plane* device.

The bipolar switching behaviour of vermiculite memristors is demonstrated in different type of VM devices . To investigate the conductance modulation with repetitive voltage sweeps for both the crossings, six triangular voltage pulses of fixed polarity were applied for crossing-1; resulting in a continuous decrease and increase in current with +1 V and −1 V pulses respectively (Fig. 4A). In contrast, crossing-2 exhibited an inverse response, with the current increasing under +1 V pulses and decreasing under -1 V pulses (Fig. 4B). To investigate the long-term *G* modification of the device, write voltage spikes of −0.55 V and +0.55 V were applied with a duration of 60 s for crossing-1 (Fig. 4C inset) and crossing-2 (Fig. 4D inset) respectively. The effect of the write pulses on the device is accessed with consecutive read pulses of ±0.15 V for 10 s ensuring no interference with the write process. The device is then reset progressively by applying the pulses of the opposite polarity for the same duration, leading to continuous decrease in the *G* states. The device regains the initial *G* state in both the crossings but attains it faster in crossing-2 (Fig. 4C, D) and hence confirming the reversible nature of the *G* state of our device. This behaviour mimics the progressive synaptic potentiation and depression representing the neuroplasticity or persistent change in synaptic strength between neurons for learning, forgetting or modification of existing information in biological brain systems. Further, the synaptic plasticity could be categorized into short-term-plasticity (*STP*) and long-term plasticity (*LTP*) based on the temporary or long-term potentiation of neural connections. We assessed the effect of a single voltage pulse of −0.55 V and +0.55 V for 120 s, applied as an external-stimuli for our device for crossing-1 and crossing-2 respectively. The *G* states of the device are assessed with oscillating read pulse train of ±0.15 V with a duration of 10 s (Fig. 4E, F inset). The decay rate is faster initially and becomes gradual with time for both the crossings (Fig. 4E, F).

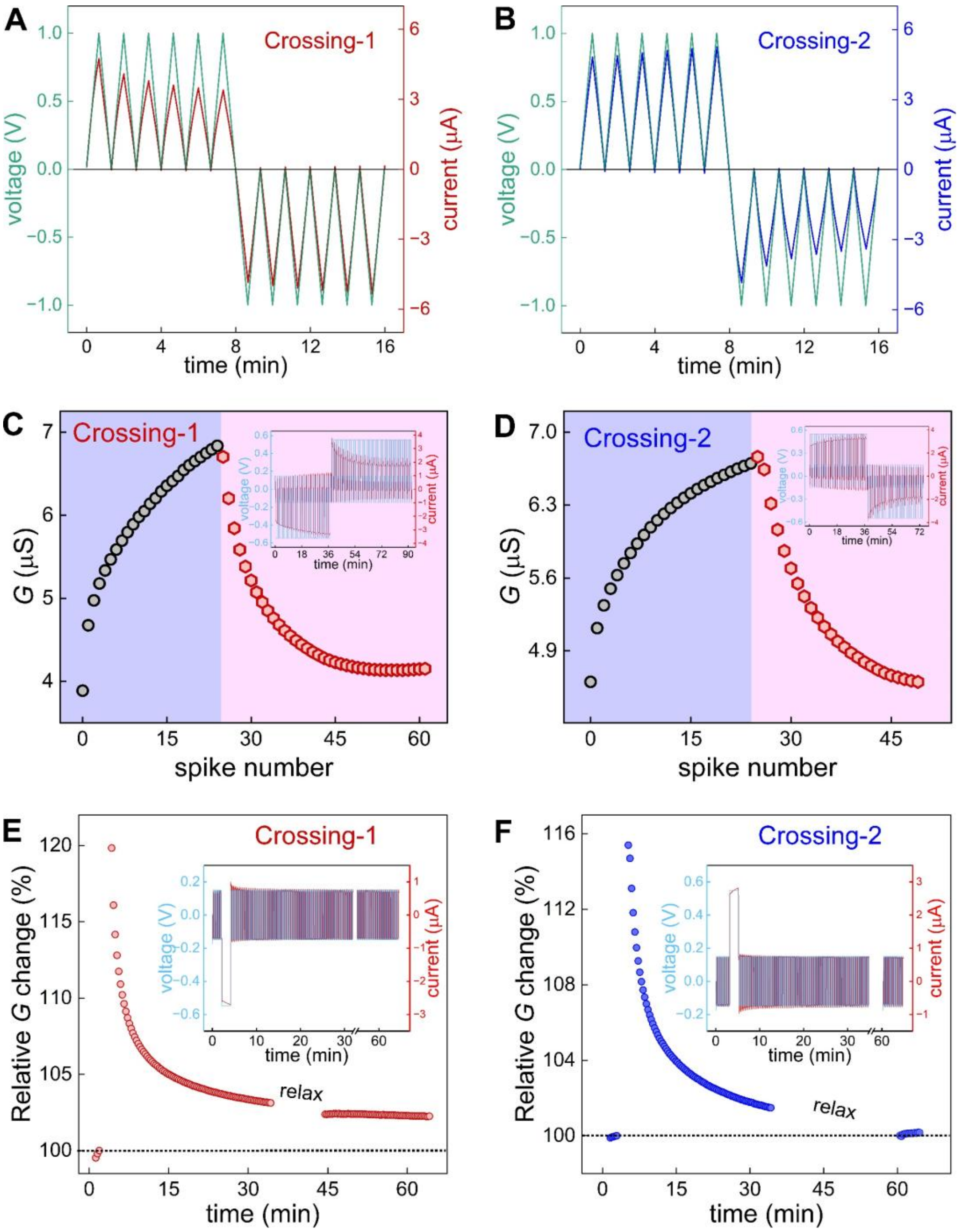

**Fig. 4. Neuromorphic mimicking of the vermiculite memristor:** Evolution of the ionic current under constant polarity voltage pulses for (**A**) crossing-1 and (**B**) crossing-2. The increase (decrease) in conductance after applying the programming pulses of −0.55 V (+0.55 V) for duration of 120 s for (**C**) crossing-1 and +0.55 V (−0.55 V) for duration of 60 s for (**D**) crossing-2. The effect of the applied write pulse is recorded with non-intrusive consecutive read pulses of ±0.15 V (10 s). A rest pulse of 0 V (5 s) is applied between the write and read pulses. Conductance changes of the VM device, in percentage of the initial conductance after applying a single write pulse of -0.55 V and +0.55 V with a duration of 120 s for (**E**) crossing-1 and (**F**) crossing-2, respectively, is accessed with the oscillating read pulses of ±0.15 V (10 s) with a rest pulse of 0 V (5 s) between them as illustrated in the inset of (E) and (F) for the respective measurements. The measurements were performed using 10 mM $AlCl_3$ electrolyte. Device V2 is used for the measurements in (**A**) and (**B**), and V4 in (**C**), (**D**), (**E**) and (**F**).

The decay rates of the VM device are estimated by fitting the exponential decay equation (Fig. S33 with the fitting):

$$G(t) = G_0 + G_{\mathrm{STP}} \exp\left(-\frac{t}{\tau_{\mathrm{STP}}}\right) + G_{\mathrm{LTP}} \exp\left(-\frac{t}{\tau_{\mathrm{LTP}}}\right) \quad (5)$$

Where, $G_0$ is the initial memory state of the device (before applying the write pulse), $G_{\mathrm{STP}}$ and $G_{\mathrm{LTP}}$ are *STP* and *LTP* related weight factors, $\tau_{\mathrm{STP}}$ and $\tau_{\mathrm{LTP}}$ are the retention time constants of *STP* and *LTP*, respectively. The fit yields $\tau_{\mathrm{STP}} \approx 95$ s and $\tau_{\mathrm{LTP}} \approx 1769$ s for crossing-1 and $\tau_{\mathrm{STP}} \approx 124$ s and $\tau_{\mathrm{LTP}} \approx 1092$ s for crossing-2 (Fig. S33). These results highlight the tunable memory behaviour of our device with higher $\tau_{\mathrm{STP}}$ for crossing-2 as compared to crossing-1 and the trend being the opposite for long term memory retention with $\tau_{\mathrm{LTP}}$ being higher for crossing-1. The retention behaviour of the VM devices could be further tuned by modulating the pulse parameters such as amplitude and duration of the write pulse (Fig. S34). To further investigate the reliability of our devices for neuromorphic computing applications, we investigated the endurance by performing successive pulse measurements (Fig. S36A) and the resistance states remain stable for more than 5 hours (133 cycles) (Fig. S36B). The reversible and programmable memory nature of our vermiculite memristors can be optimised based on the applications. The non-linearity observed in our measurements can be utilised for temporal information processing.[38] The observed short-term plasticity time constant ($\tau_{STP}{\sim}100s$, Fig. S33) extends the usable temporal window, enabling longer mask sequences before saturation.[39] This is advantageous for low-bandwidth tasks, as it allows more virtual nodes to evolve within a single memory window while preserving nonlinear fading memory and hence making the device beneficial for reservoir computing.

**Conclusions**

In summary, we report the tunable memory states of the *out-of-plane* VM channels with negative surface charge layers. The VM device is designed in a way that it exhibits switching of the memory between crossing-1 and crossing-2; independent of the electrolyte type and the concentration. Further, the inherent memristive characteristics of the VM clay were investigated by changing the VM channel configuration from *out-of-plane* to *in-plane* resulting in the diminished memory effects. Our neuromorphic studies highlight the reversible nature of easily fabricated and scalable VM memristors, enabling programmable *STP* and *LTP* states for diverse applications.

**Acknowledgements**

B.R. acknowledges funding from the Royal Society University Research Fellowship URF\R\231008, Philip Leverhulme Prize PLP-2021-262, European Union's H2020 Framework Programme\ERC Starting Grant 852674 AngstroCAP, EPSRC new horizons grant EP/X019225/1, EPSRC strategic equipment grant EP/W006502/1. S.D.G acknowledges financial support provided by Indian Institute of Technology Kharagpur (IIT-KGP), India [Sanction Letter no.: IIT/SRIC/ATDC/CEM/2013-14/118]. S.Y. acknowledges IIT-KGP and University of Manchester (UoM) for PhD fellowship. R.K.G. acknowledges Research Collaboration Grant from UoM.

**Author Contributions**

B.R. designed and directed the project. B.R. and S.D.G provided supervision for S.Y. S.Y. fabricated and characterised the vermiculite membranes and devices. S.Y. and R.K.G performed measurements, analyzed and provided all the data. A.L. performed the COMSOL simulation and provided corresponding data. S.V.P. provided inputs in neuromorphic measurements. A.B. fabricated the ultramicrotomed VM device. B.R., S.Y. and R.K.G. wrote the manuscript with inputs from A.L and S.V.P. All the authors contributed to discussions.

**Data availability**

The data of this study are available in https://figshare.com/s/0f822a6589bcfd0bd45e

**Declaration of Competing Interest**

The authors declare no known competing interests.